# Orbit topology analysed from π phase shift of magnetic quantum oscillations in three-dimensional Dirac semimetal

Sang-Eon Lee,[1] Myeong-jun Oh,[2] Sanghyun Ji,[1] Jinsu Kim,[1] Jin Hyun Jun,[1] Woun Kang,[3] Younjung Jo,[2,*] Myung-Hwa Jung[1,*]

[1]*Department of Physics, Sogang University, Seoul 04107, Korea*
[2]*Department of Physics, Kyungpook National University, Daegu 41566, Korea*
[3]*Department of Physics, Ewha Womans University, Seoul 03760, Korea*

## Abstract

With the emergence of Dirac fermion physics in the field of condensed matter, magnetic quantum oscillations (MQOs) have been used to discern the topology of orbits in Dirac materials. However, many previous researchers have relied on the single-orbit Lifshiftz–Kosevich formula, which overlooks the significant effect of degenerate orbits on MQOs. Since the single-orbit LK formula is valid for massless Dirac semimetals with small cyclotron masses, it is imperative to generalize the method applicable to a wide range of Dirac semimetals, whether massless or massive. This report demonstrates how spin-degenerate orbits affect the phases in MQOs of three-dimensional massive Dirac semimetal, $NbSb_2$. With varying the direction of the magnetic field, an abrupt π phase shift is observed due to the interference between the spin-degenerate orbits. We investigate the effect of cyclotron mass on the π phase shift and verify its close relation to the phase from the Zeeman coupling. We find that the π phase shift occurs when the cyclotron mass is 1/2 of the electron mass, indicating the effective spin gyromagnetic ratio is $g_s = 2$. Our approach is not only useful for analysing MQOs of massless Dirac semimetals with a small cyclotron mass, but also can be used for MQOs in massive Dirac materials with degenerate orbits, especially in topological materials with a sufficiently large cyclotron mass. Furthermore, this method provides a useful way to estimate the precise $g_s$ value of the material.

# I. Introduction

Since the observation of the non-trivial π Berry phase obtained from magnetic quantum oscillations (MQOs) in the typical Dirac material of graphene (1), the π phase in MQOs has become a signature of Dirac fermions. Subsequently, researchers have revealed Dirac fermions from MQOs in topological insulators (2–5) with Dirac cones in their surface electronic structures (6–11). With the emergence of Dirac and Weyl semimetals with bulk electronic states effectively representing Dirac and Weyl fermions (12–17), the π phase in MQOs has been widely used to detect the topology of orbits (18–23). For a variety of topological materials, many researchers have used a single-orbit Lifshitz Kosevich (LK) formula, which represents the oscillation of conductivity or magnetization induced by consecutively passing the quantized orbit of Fermi surface in a non-degenerate band with reduced amplitude of MQOs due to finite temperature and relaxation time. In the single-orbit LK formula, the total oscillation is expressed by the Fourier series. Thereby, the fundamental harmonics include $\cos(2\pi F/B - \pi + \lambda + \delta)$, where $F$ is the frequency of the MQOs, $B$ is the magnetic field, $\lambda$ is the phase considered to represent the Berry phase, and $\delta$ is the phase associated with the Fermi-surface curvature.

For the surface states of a topological insulator, $\lambda$ is directly related to the observed phase $\Theta$. However, in the case of spin-degenerate systems such as graphene and Dirac semimetal, the superposed oscillations force the relation between $\Theta$ and $\lambda$ to be $\Theta = \frac{\pi}{2}[1 - \text{sign}(\cos\lambda)]$ (24). The amplitude of superposed oscillations is proportional to $|\cos\lambda|$. This effect of superposed oscillations is governed by how the spin-up and spin-down oscillations interfere, which is related to the relative ratio between the Landau splitting and the spin splitting. In other words, the amplitude and the phase of MQOs vary depending on the direction of the magnetic field in anisotropic materials. The change in oscillation amplitudes has been extensively reported in quasi-two-dimensional materials such as semiconductor heterojunctions (25, 26) and organic superconductors (27, 28). However, the orbit topology through the phase analysis has rarely been addressed in these materials. Moreover, little consideration has been given to the relationship between the amplitude and phase of MQOs in Dirac materials.

The $\lambda$ associated with orbit topology is given by $\lambda = \varphi_B + \varphi_R + \varphi_Z$. Here, the orbital contribution is $\varphi_B + \varphi_R$, where $\varphi_B$ and $\varphi_R$ are related to the Berry phase and the orbital magnetization, respectively. The spin contribution is $\varphi_Z$, which comes from the spin magnetization related to the Zeeman coupling. There have been many studies analyzing MQOs, considering that the observed phase is directly related to the $\varphi_B$. In most cases, for massless Dirac semimetals such as graphene (1), $Cd_3As_2$ (18), $Na_3Bi$ (29), and $ZrTe_5$ (30), the orbit topology could be determined only from $\varphi_B$, ignoring other phases. The $\phi_B$ obtained from the oscillation of a closed obit is topologically discretized to $\pi$ or 0, depending on whether the orbit encloses a band touching point or not (31). On the other hand, in the case of massive Dirac semimetals, $\varphi_B$ is not an appropriate phase for analyzing the orbit topology because $\varphi_B$ is not discretized and can change continuously depending on the gap size. Instead, $\varphi_B + \varphi_R$ is a more appropriate phase than $\varphi_B$ since $\varphi_B + \varphi_R$ is discretized to $\pi$ or 0, regardless of whether the system is massless or massive (24,29). In order to obtain the orbit topology, it is essential to separate $\phi_Z$ in $\lambda$. Since $\phi_Z$ is proportional to $m_c/m_e$, where $m_c$ is the cyclotron mass and $m_e$ is the electron mass, for massive Dirac semimetals with a sufficiently large $m_c$, the orbit topology can be determined by evaluating the exact value of $\phi_Z$.

Recently, some studies have been conducted on the incorporation of spin-degenerate orbits in a Dirac material. For instance, a $\pi$ phase shift in MQOs that is dependent on the angle between the crystal axis and the magnetic field has been observed in $ZrTe_5$, a three-dimensional Dirac semimetal (30, 32–34). The $\pi$ shift was interpreted as the angle-dependent spin reduction factor (30), implying the effect of $\phi_Z$. Since $ZrTe_5$ is a quasi-two-dimensional material (35), it has strong anisotropy, which may induce angle-dependent $\phi_Z$. Therefore, it is not clear whether the $\pi$ phase shift can be observed in three-dimensional (not quasi-two-dimensional) Dirac semimetals. The origin of the $\pi$ phase shift in $ZrTe_5$ has been proposed to be associated with the Zeeman coupling, but the phase effect of orbital magnetization has not been considered for the MQO analyses.

Since most of the MQO analyses have focused on massless Dirac semimetals and have not been applied well to experimental studies, in our study, we attempted to analyze the orbit topology in a massive Dirac semimetal with a more general concept. For a more general and complete understanding of the phases of MQOs, we studied MQOs of the massive Dirac semimetal $NbSb_2$, which is an electron-hole compensated semimetal with multiple band-touching points without spin-orbit coupling. The spin-orbit coupling splits the band touching points and opens the gaps (36–38). The massive Dirac structure and the large cyclotron mass (~$0.5m_e$) require the integration of $\phi_Z$ and $\phi_R$ as well as $\phi_B$, which implies that $NbSb_2$ is a good platform to adapt the LK formula with a more general concept of orbit topology with spin-degnerate orbits. Our work paves the way for defining orbit topologies in a variety of materials through MQOs.

## II. Preliminary theory

Considering the superposed oscillation with spin-degenerate orbits, the form of the LK formula should be $\cos(2\pi F/B - \pi + \lambda + \delta) + \cos(2\pi F/B - \pi - \lambda + \delta) = 2|\cos\lambda|\cos(2\pi F/B - \pi + \Theta + \delta)$, where $\Theta = \frac{\pi}{2}[1 - \text{sign}(\cos\lambda)]$, as shown in Ref. (24). Incorporating the reduction factor and higher-order harmonics, the ratio of the oscillatory part of the conductivity $\Delta\sigma_{xx}$ to the non-oscillatory part $\sigma_{xx}^{no}$, $\Delta\sigma_{xx}/\sigma_{xx}^{no}$, for a three-dimensional Dirac semimetal is (19, 24, 39, 40)

$$\frac{\Delta\sigma_{xx}}{\sigma_{xx}^{no}} \approx \sum_{p=1}^{\infty} A_\sigma \sqrt{B/p} R_T R_D |\cos p\lambda| \cos\left[ p\left(\frac{2\pi F}{B} - \pi\right) + \Theta + \delta \right], \tag{1}$$

where $A_\sigma$ is a dimensional constant, $p$ is the harmonic order, $R_T = \dfrac{2\pi^2 p k_B T/\hbar\omega_c}{\sinh(2\pi^2 p k_B T/\hbar\omega_c)}$ is the temperature reduction factor, and $R_D = \exp(-\pi p/\tau\omega_c)$ is the Dingle reduction factor.

## III. Results and discussion

We measured the magnetoresistance (MR) of NbSb$_2$ with varying the angle $\theta$ between the current direction (b-axis) and the direction of the magnetic field (see Fig. 1*A, Inset*). The elongated direction of the sample is along the b-axis (36, 37). Fig. 1*A* shows the MR curves of NbSb$_2$ from $\theta = -90°$ to $\theta = 90°$. The MR gradually increases with increasing $|\theta|$. When the magnetic field is applied perpendicular to the current direction ($\theta = \pm 90°$), an extremely large MR is evident (~1.8×10$^6$% at 14 T and 1.5 K), which exceeds previously reported MR values (36, 37). In semimetals such as NbSb$_2$, the large MR is attributed to the almost perfectly compensated carrier density and high crystal quality (37, 41–44). To estimate the carrier density, we measured the Hall resistance of NbSb$_2$ and fit the data with a two-carrier model. Consequently, we obtain an almost compensated carrier density (Supplementary Section 1 and Supplementary Fig. S3). The compensated carrier density is well matched with the previous density functional theory calculation results (38). Furthermore, the high crystal quality is manifested in the high residual resistivity ratio (RRR) of ~520 (Supplementary Fig. S3), which is higher than the values reported previously (36, 37).

SdH oscillations are clearly observable in the MR data. We analysed these oscillations using a fast Fourier transform (FFT). Fig. 1*B* shows a typical FFT spectrum at $\theta = 60°$. We define the peaks in Fig. 1*B* as α and β oscillations and their second harmonics as 2α and 2β. The observed frequencies are similar to those reported previously (36, 37) (Supplementary Section 2 and Supplementary Fig. S4). The colormap in Fig. 1*C* represents the angle-dependent FFT amplitude and the symbols indicate the frequencies of the peaks selected in the FFT spectrum. It includes major α (black closed circles) and β (red closed circles) oscillations and their second harmonics (black and red open circles) as well as the minor ζ (magenta stars), η (cyan rhombuses), and κ (blue triangle) oscillations. The overall behaviour of the angle-dependent frequencies is symmetric to $\theta = 0°$, where the applied magnetic field is along the two-fold symmetric axis of the crystal (b-axis). The α and β oscillations span all angles and monotonically vary with the angle, whereas the ζ, η, and κ oscillations are observable for only a small range of angles.

In order to analyse the MQOs in detail, we convert the conductivity form in equation (1) into the form of resistivity using the recent transport theory for MQOs (45) (Supplementary Section 3). We obtain the form of inverse resistivity, $\Delta \rho_{xx}^{-1} / \rho_{xx}^{no-1} \approx \sum_i f_i \varepsilon_i$ by introducing a correction factor $f_i$ which is added to the oscillatory part of the conductivity from the *i*th carrier $\varepsilon_i$. Consequently, we obtain

$$\frac{\Delta \rho_{xx}^{-1}}{\rho_{xx}^{no-1}} \approx \sum_i \sum_{p=1}^{\infty} A_{\rho,i} \sqrt{B/p} R_{T,i} R_{D,i} |\cos p\lambda_i| |f_i| \cos\left[p\left(\frac{2\pi F_i}{B} - \pi\right) + \Theta_i + \delta_i + \varphi_i\right], \qquad (2)$$

where $A_{\rho,i}$ is a dimensional constant. We introduce the phase $\varphi_i = \frac{\pi}{2}[1 - \text{sign}(f_i)]$. If $f_i$ is positive (negative), the phases of inverse resistivity and conductivity are the same (differ by π). Furthermore, the shape of the oscillations can be modified if $f_i$ changes the sign with varying the magnetic field.

From the polynomial fits of the measured MR curves, we obtain the non-oscillatory and oscillatory parts of the resistance and convert them into the form of inverse resistivity, $\Delta \rho_{xx}^{-1} / \rho_{xx}^{no-1}$. We select the frequency range encompassing the α and β oscillations (red shaded area in Fig. 1B) and filter out all other frequencies. The upper panel of Fig. 1D shows the filtered data and the fitted curve obtained from equation (2) (here, we set $|f_i| = 1$). The agreement between them justifies applying the LK theory to the filtered data. The α and β oscillations are well reconstructed separately from the fitted curve, as shown in the lower panel of Fig. 1D.

Now, we discuss the amplitude and the phase obtained from the above analyses. It is reasoanble to choose a set of frequencies if the peaks are well distinguished from other freuqencies and the frequency changes continuously depending on the angle of the magnetic field. We selected a reliable angle range and a set of α frequency for the SdH oscillations to apply the band-pass filter. In Fig. 2A, we plot the FFT amplitudes of the α oscillations and the 2α oscillations, A(α) and A(2α), respectively. For most angles, A(α) is much larger than A(2α), except at $\theta = \pm 40°$ where A(α) becomes very small, even smaller than A(2α). The phases of the α oscillations are obtained by drawing the Landau-level fan diagram (Supplementary Section 4 and Supplementary Fig. S5, S6). As is conventionally done (1, 21, 30), we divide the obtained phase of the α oscillations by $2\pi$, which is written as $\gamma_\alpha = (-\pi + \Theta_\alpha \pm \delta_\alpha + \varphi_\alpha)/2\pi$. Fig. 2B represents A(2α)/A(α) and $\gamma_\alpha$ as a function of $\theta$. Interestingly, $\gamma_\alpha$ abruptly changes by 0.5 (corresponding to $\pi$ phase shift) near $\theta = \pm 40°$, where A(2α)/A(α) increases. Here, we do not present an analysis of the α oscillations at $\theta = -10°–10°$, where the η oscillations are close to the α oscillations. In equation (2), the related amplitude factors are $|\cos p\lambda_\alpha|$ and $|f_\alpha|$. However, the behaviour of the amplitudes at $\theta = \pm 40°$ is different between the fundamental harmonics ($p = 1$) and the second harmonics ($p = 2$); hence, the $p$-dependent amplitude factor $|\cos p\lambda_\alpha|$ should be related. Since the factor is the cosine function, the observed large A(2α)/A(α) is attributable to the fact that $\lambda_\alpha$ is close to $(n + 1/2)\pi$, where $n$ is an integer. The phase factor related to $\lambda_\alpha$ is

$$\Theta_\alpha = \frac{\pi}{2}[1 - \mathrm{sign}(\cos \lambda_\alpha)],$$ and one can recognize that when $\lambda_\alpha$ is close to $(n + 1/2)\pi$, $\Theta_\alpha$ is shifted by $\pi$. There may be alternative explanations for the $\pi$ phase shift with respect to the relaxiation time. When measuring MR while changing the angle between the current and the magnetic field, there can be a phase difference of $\pi$ between the longitudinal and transverse directions (45). In that case, the phase of MQOs should continuously change with the direction of the magnetic field. However, in our results, the phase of the α oscillation shows step-like change at the angle where the $\pi$ phase shift occurs. Furthermore, the ratio of 2α to α amplitude, A(2α)/A(α) rapidly increases. With respect to these facts, our observation is mainly explained by the interference effect between spin-

degenerate orbits, and other factors of phase shift due to the sign change in the oscillatory part can be excluded.

The α oscillations at $\theta = \pm 90°$ as shown in Fig. 2C, an unconventional node appears at $1/B \approx 0.12$. In addition, the phase changes continuously depending on the magnetic field. The phase difference between the oscillation phase evaluated in the high field region ($\gamma_{high}$) and the low field region ($\gamma_{low}$) is substantial as 0.280. The node-like shape and field-dependent phase can be explained by $f_i$. Figure 2D is the result of simulating the sign of $f_i$ that is closely related to the phase, assuming that there are two carrier pockets. The sign of $f_i$ is the function of carrier density ratio $n_2/n_1$ and inverse magnetic field $1/B$. The mobilities of the carrier pockets are fixed to the values of NbSb$_2$, $\mu_1 = 1.68$ m$^2$/Vs and $\mu_2 = 2.42$ m$^2$/Vs (Supplementary Section 1). The yellow (purple) area represents the positive (negative) sign of $f_1$, which means that the phases of inverse resistivity and conductivity are the same (differ by π). The yellow region may appear when the magnetic field is weak and the carrier density is compensated, i.e. $n_2 = -n_1$. In addition, its origin is related to the Hall conductivity $\sigma_{xy}$ of the conductivity-resistivity transformation tensor $\rho_{xx} = \frac{\sigma_{xx}}{\sigma_{xx}^2 + \sigma_{xy}^2}$. If $\sigma_{xy}$ is small (e.g., $n_2 = -n_1$), the sign of $f_1$ is positive (yellow region); otherwise (e.g., $n_2/n_1 > 0$) the sign of $f_1$ is negative (purple region). If $f_1$ changes the sign depending on the magnetic field, a node will appear at the sign change point, where a phase change occurs as well. We reconstructed the α oscillations with the sign change of $f_1$. We chose the parameters to be $m_c = 0.92 m_e$, $\tau = 10$ ps, $F = 300$ T, $\lambda = 0$, and $n_2/n_1 = -1.13$ (red line in Fig. 2D) and plotted the oscillations obtained using equation (2) as a function of $1/B$ in Fig. 2E ($m_c = 0.92 m_e$ was taken from Fig. S10). A node appears at $1/B \approx 0.12$, which is similar to the node position of the α oscillations at $\theta = 90°$. It is noteworthy that the node-like shape and field-dependent phase can be evidence of a compensated carrier density, since $f_i$ changes signs only when the charge carrier density is nearly compensated if the magnetic field is sufficiently strong. The slightly different shapes in Figs. 2C and 2E are attributable to the fact that we only assumed two carriers and used an FFT band-pass filter that can distort the oscillation shape.

In addition to the SdH oscillations, we measured the dHvA oscillations of NbSb$_2$ to check how well the oscillations match each other and how much the above analysis is reliable. Fig. 3A shows the FFT contour map of the dHvA oscillations, and the oscillations are defined as done in Fig. 1C. The small difference of the frequency between the SdH and dHvA oscillations are most likely due to a small misalignment in the measurement. The oscillatory part of the magnetization $\Delta M$ is given by (19, 24, 40)

$$\Delta M \approx \sum_i \sum_{p=1}^{\infty} -A_{M,i}\sqrt{B/p}R_{T,i}R_{D,i}|\cos p\lambda_i|\sin\left[p\left(\frac{2\pi F_i}{B}-\pi\right)+\Theta_i+\delta_i\right], \tag{3}$$

where $A_{M,i}$ is a dimensional constant. The phase $\Theta_i$ of the dHvA oscillations is obtained, and the phase parameter is defined as $\gamma'_i = (-\pi + \Theta_i \pm \delta_i)/2\pi$ (Supplementary Section 4 and Supplementary Fig. S7, S8). We compare the phase $\gamma'_\beta$ to the FFT amplitude ratio A(2β)/A(β) in Fig. 3B. Similar to the SdH oscillations in Fig. 2B, the π phase shift occurs near $\theta = 40°$, where A(2β)/A(β) increases, indicating that $\lambda_\beta$ produces the π phase shift. As aforementioned, this phase shift is not necessary to be the Berry phase but is composed of three phases; $\lambda = \phi_B + \phi_R + \phi_Z$ where $\phi_B$, $\phi_R$, and $\phi_Z$ are related to the Berry phase, the orbital magnetization, and the Zeeman coupling, respectively.

Next, we discuss the origin of the π phase shift. For this purpose, we assume that the predominant properties of the orbit can be determined by considering only two bands (called as two-band approximation). Previous theoretical works based on this assumption have shown that $\phi_B + \phi_R$ is approximately fixed to either 0 or π (24, 31) for massive Dirac dispersion. Since the value of $\phi_B + \phi_R$ indicates the number of band touching points that the orbit encloses, when the number is even (odd), the value is zero (π). Therefore, we obtain $\lambda \approx \phi_Z$ or $\pi + \phi_Z$, so that the only possible origin of the amplitude-dependent π phase shift in our case is $\phi_Z$, resulting that the π phase shift occurs when $\phi_Z = (n + 1/2)\pi$ (see Supplementary Movie S1).

The π phase shift is related to the cyclotron mass, which enables determining the orbit topology. We measured the temperature-dependent dHvA oscillations at some specific angles of $\theta = 0°$, 40°, and 90° to estimate the cyclotron mass. The temperature-dependent FFT amplitudes of the oscillations at $\theta = 40°$ is plotted in Fig. 4A (Supplementary Fig. S9 presents the data corresponding to the other angles), where the ζ, α, β, and 2β oscillations are shown. It is clearly seen that the amplitude decreases with increasing temperature. Notably, the amplitude of the 2β oscillations is larger than that of the β oscillations, indicating the π phase shift at this angle. Fig. 4B depicts the temperature-dependent amplitudes of the β oscillations for $\theta = 0°$, 40°, and 90°, which are well fitted by using the temperature reduction factor $R_T$. The cyclotron mass is estimated to be $m_c = 0.524 \pm 0.002$ $m_e$, $0.475 \pm 0.013$ $m_e$, and $0.374 \pm 0.006$ $m_e$, respectively. Since $\Delta M$ is smoothly varying with $\theta$, it seems that the $\pi$ phase shift occurring near $\theta = 40°$ is related to $m_c \approx 0.5 m_e$.

To understand the significance of the $\pi$ phase shift when $m_c \approx 0.5 m_e$, we should look at $\phi_Z$. Roughly speaking, $\phi_Z$ indicates the relative size of the Zeeman energy splitting of an orbit compared to the Landau level splitting. Since the Landau level splitting decreases with increasing the cyclotron mass, $\phi_Z$ is proportional to the cyclotron mass; $\phi_Z = \pi \frac{g_s m_c}{2 m_e}$, where $g_s$ is the effective gyromagnetic ratio of the spin (31, 46). In general, the $g_s$ is not exactly 2 because of the effect of the spin-orbit coupling. However, for the β oscillations, since the frequency of the MQOs is relatively high between 400 and 700 T, one can suppose that the energy difference between the Fermi level and the band edge is large. Thus, the kinetic energy of the electrons at the Fermi level is high and will dominate the spin-orbit coupling energy, which corresponds to $g_s \approx 2$, leading to $\phi_Z = \pi \frac{m_c}{m_e}$. With this expression, $\lambda$ can be written as a function of $m_c$; $\lambda = \pi \frac{m_c}{m_e}$ for a trivial orbit and $\lambda = \pi \left(1 + \frac{m_c}{m_e}\right)$ for a non-trivial orbit.

Finally, integrating the above statements, we discuss the orbit topology of the β oscillations. At $\theta = 90°$, from $m_c = 0.37 m_e$ we obtain $0 < \phi_Z < \frac{1}{2}\pi$. The fact that the only possible values of $\phi_B + \phi_R$ are 0 or $\pi$ results in $0 < \lambda < \frac{1}{2}\pi$ for $\phi_B + \phi_R = 0$ and $\pi < \lambda < \frac{3}{2}\pi$ for $\phi_B + \phi_R = \pi$. In this case, from $\Theta = \frac{\pi}{2}[1 - \text{sign}(\cos \lambda)]$, $\Theta = 0$ for $\phi_B + \phi_R = 0$ and $\Theta = \pi$ for $\phi_B + \phi_R = \pi$, which means that $\Theta$ and $\phi_B + \phi_R$ are the same, and we can find the orbit topology based on the measured value of $\Theta$. From $\gamma_\beta = (-\pi + \Theta_\beta + \delta_\beta)/2\pi \approx 0.35$ for the β oscillations at $\theta = 90°$ (Fig. 3B), $\Theta_\beta \approx 1.7\pi - \delta_\beta$ (mod $2\pi$) $= -0.3\pi - \delta_\beta$. Plugging in $\delta_\beta \approx -0.25\pi$ (19, 40), from $\Theta \approx 0$, we obtain $\phi_B + \phi_R \approx 0$, which means that the orbit producing β oscillations encloses an even number of band-touching points.

Here, we assume β oscillations come from the ellipsoidal pockets, indicating $\delta_\beta \approx -0.25\pi$, since the angle-dependent frequencies of β oscillations are monotonically varied. All the above discussion can be reproduced from the second harmonic of the β oscillations, 2β with twice the cyclotron mass, and the results are consistent (see Supplementary Section 5 and Fig. S11). It is noteworthy that the π phase shift will not occur at $m_c \approx 0.5m_e$ if one of the assumptions (the two-band approximation or $g_s \approx 2$) is not valid. We found this situation to be the case for the ζ oscillations, and Supplementary Section 5 and Fig. S11 provides the related discussion.

The results demonstrate that our method is suitable for analysing the MQOs in Dirac semimetals. Unlike the phase of the MQOs in massless Dirac semimetals or topological insulators, which is properly analysed with the single-orbit LK thoery, the phase of the MQOs in massive Dirac semimetals must be analysed more carefully because of the spin-degnerate orbits and the complex phase contributions. Equations (1)-(3) constitute the most suitable means for analysing the MQOs in Dirac semimetals. We measured the SdH and dHvA oscillations of the massive Dirac semimetal $NbSb_2$ and analysed the MQOs using the aforementioned equations. Both types of oscillations clearly showed π phase shifts, which are the manifestations of spin-degenerate orbits. We verified that the origin of the π phase shift observed in the dHvA oscillations is the Zeeman coupling, not the Berry phase. Most importantly, we determined the orbit topology using the method that considered the spin-degenerate states and the phase from orbital magnetization, which completely differs from the conventional method used for massless Dirac semimetlas or topological insulators. Our approach will not only be useful to analyse the MQOs of Dirac semimetals with the spin degeneracy, but also provide perspectives for analysing the MQOs of various materials with other types of degeneracy.

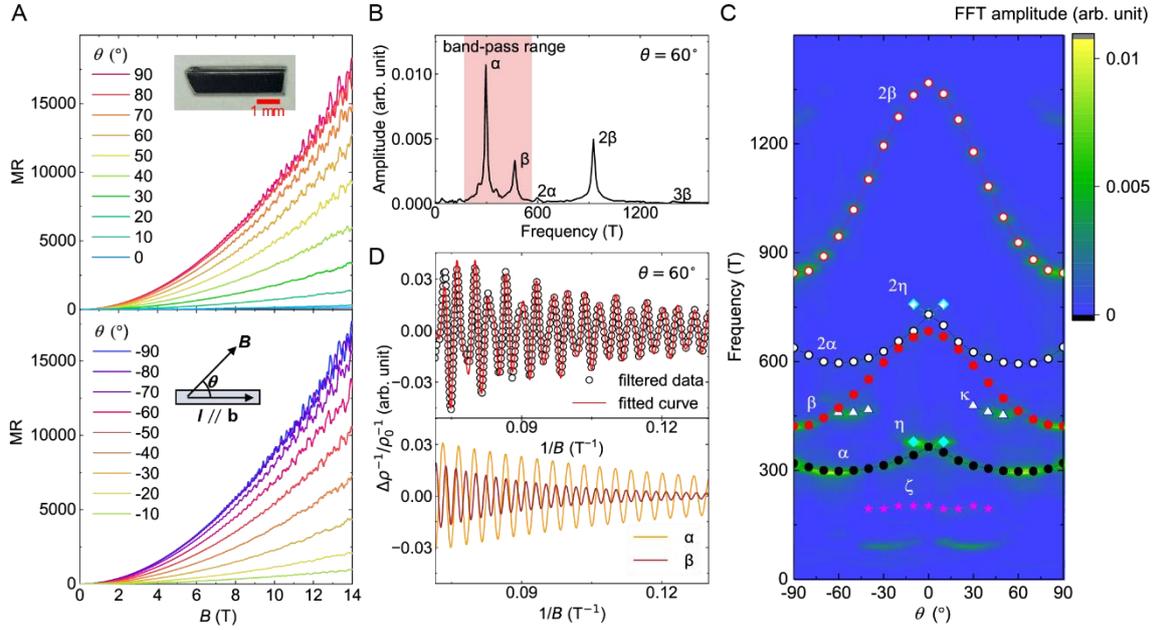

**Figure 1.** MR and SdH oscillations of NbSb$_2$. (*A*) MR of NbSb$_2$ at $\theta$ = -90°–90°. The current direction is along the crystal b-axis, and $\theta$ is defined as the angle between the current and magnetic field directions. (*Insets*) Picture of the crystal (upper panel) and the experimental configuration (lower panel). The extremely large MR of 1.8×10$^6$% (at 14 T and 1.5 K) is observable when the current is perpendicular to the magnetic field ($\theta$ = ±90°). (*B*) FFT spectrum of NbSb$_2$ at $\theta$ = 60°. The $\alpha$ and $\beta$ oscillations, second harmonic 2$\alpha$ and 2$\beta$ oscillations, and third harmonic 3$\beta$ oscillations are defined. The red shaded area displays the FFT band-pass filter range used to filter the $\alpha$ and $\beta$ oscillations. (*C*) Angle-dependent FFT frequencies. The colormap represents the FFT amplitude, and the symbols represent the frequencies of the FFT peaks. $\alpha$, $\beta$, $\zeta$, $\eta$, and $\kappa$ oscillations are identified (*D*) Filtered data the band-pass range in *B* and fitted curve with equation (2). The agreement between filtered data and fitted curve shows that it is appropriate to apply the LK theory (upper panel). Reconstructed $\alpha$ and $\beta$ oscillations obtained from the fitted curve in the upper panel (lower panel).

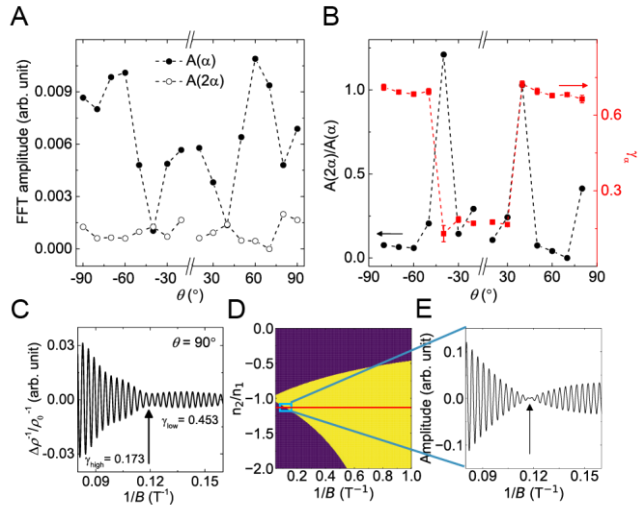

**Figure 2.** Angle-dependent α oscillations in the SdH oscillations of NbSb$_2$. (*A*) Amplitude of α oscillations A(α) and second harmonic 2α oscillations A(2α). A(α) is small at $\theta = \pm 40°$, even smaller than A(2α). (*B*) Amplitude ratio A(2α)/A(α) and phase of α oscillations $\gamma_\alpha$. $\gamma_\alpha$ changes rapidly by 0.5 (corresponding to π phase shift) near $\theta = \pm 40°$, where the value of A(2α)/A(α) increases. The coincidence of large A(2α)/A(α) and π phase shift indicates that the π phase shift stems from the angle-dependent interference between spin-degenerate orbits. The standard errors in the Landau fan diagram fitting are displayed, but are not visible because they are smaller than the symbol size (~0.03). (*C*) Node shaped α oscillation at $\theta = 90°$, and the arrow indicates the node position. The phases evaluated in high field region ($\gamma_{high}$) and low field region ($\gamma_{low}$) differ by 0.280. (*D*) Sign of $f_1$ simulated by equation (S7) as a function of carrier density ratio $n_2/n_1$ and inverse magnetic field $1/B$. The yellow (purple) region indicates the positive (negative) sign. The mobility parameters $\mu_1 = 1.68$ m$^2$/Vs and $\mu_2 = 2.42$ m$^2$/Vs were selected based on the two-carrier fitting (supplementary Fig. S2). (*E*) α oscillation obtained by simulating equation (2) using parameters from the area marked by sky-blue square in *D*, where $m_c = 0.92 m_e$, $\tau = 10$ ps, $F = 300$ T, $\lambda = 0$, and $n_2/n_1 = -1.13$ The position of the node indicated by the arrow is similar to that of *C*.

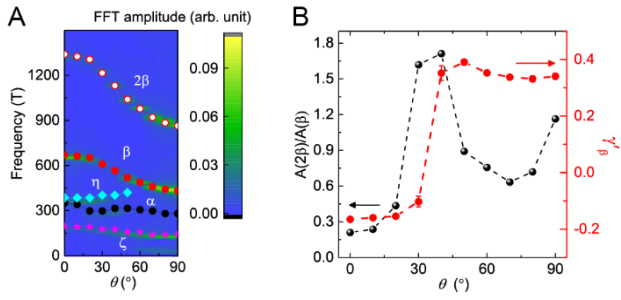

**Figure 3.** Angle-dependent dHvA oscillations in NbSb$_2$. (*A*) Angle-dependent FFT amplitudes and frequencies of dHvA oscillations. α, β, ζ, and η oscillations are identified. (*B*) Amplitude ratio A(2β)/A(β) and phase of β oscillations $\gamma_\beta$. $\gamma_\beta$ changes rapidly by 0.5 (corresponding to π phase shift) near $\theta = 40°$, where the value of A(2β)/A(β) increases. The standard errors in the Landau fan diagram fitting are displayed, but are not visible because they are smaller than the symbol size (~0.03).

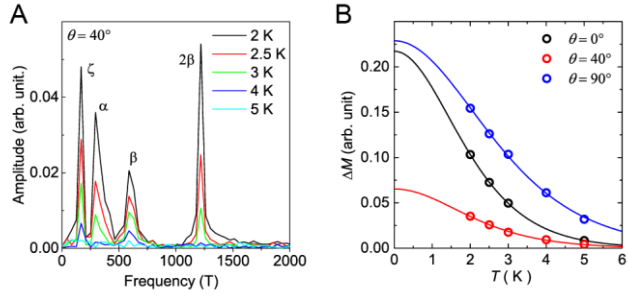

**Figure 4.** Temperature-dependent dHvA oscillations of NbSb$_2$. (*A*) FFT amplitude of dHvA oscillations at $\theta$ = 40° for various temperatures. (*B*) Temperature-dependent FFT amplitudes of β oscillations at $\theta$ = 0°, 40°, and 90° (open symbols). The amplitudes are fitted with temperature reduction factor $R_T$ (solid lines) for the cyclotron mass evaluation. The cyclotron mass is evaluated to be $m_c$ = 0.52$m_e$, 0.48$m_e$, and 0.37$m_e$ at $\theta$ = 0°, 40°, and 90°, respectively.